\documentclass[
superscriptaddress,
aip,apl,
reprint,
twocolumn,
amsmath,amssymb,showpacs
]{revtex4-2}
\usepackage{titlesec}
\usepackage{newtxtext,newtxmath}
\usepackage{physics}
\usepackage{graphicx}
\usepackage{dsfont}
\usepackage{xcolor}
\usepackage{mathbbol}
\definecolor{urlc}{RGB}{58,105,157}
\usepackage[
colorlinks=true,
urlcolor=urlc,
linkcolor=urlc,
citecolor=urlc
]{hyperref}
\titleformat{\section}{\large\sffamily\bfseries}{\thesection}{}{}
\titleformat{\subsection}{\sffamily\bfseries}{\thesubsection}{}{}
\titlespacing*{\section}{0pt}{3ex}{0ex}
\titlespacing*{\subsection}{0pt}{2ex}{0ex}

\usepackage{algorithm}
\usepackage{algpseudocode}

\begin{document}
\widetext

\title{Iterative Gradient Ascent Pulse Engineering algorithm for quantum optimal control}

\author{Yuquan Chen}
\thanks{These authors contribute equally}
\affiliation{
Hefei National Laboratory for Physical Sciences at the Microscale and Department of Modern Physics, University of Science and Technology of China, Hefei 230026, China}
\affiliation{
CAS Key Laboratory of Microscale Magnetic Resonance, University of Science and Technology of China, Hefei 230026, China}
\affiliation{
Synergetic Innovation Center of Quantum Information and Quantum Physics, University of Science and Technology of China, Hefei 230026, China}

\author{Yajie Hao}
\thanks{These authors contribute equally}
\affiliation{
Department of Physics, 
Southern University of Science and Technology, Shenzhen 518055, China.}
\affiliation{
Shenzhen Institute for Quantum Science and Engineering, 
Southern University of Science and Technology, Shenzhen 518055, China.}

\author{Ze Wu}
\affiliation{
Hefei National Laboratory for Physical Sciences at the Microscale and Department of Modern Physics, University of Science and Technology of China, Hefei 230026, China}
\affiliation{
CAS Key Laboratory of Microscale Magnetic Resonance, University of Science and Technology of China, Hefei 230026, China}
\affiliation{
Synergetic Innovation Center of Quantum Information and Quantum Physics, University of Science and Technology of China, Hefei 230026, China}

\author{Bi-Ying Wang}
\affiliation{Central Research Institute, Huawei Technologies, Shenzhen, 518129, China}

\author{Ran Liu}
\affiliation{
Hefei National Laboratory for Physical Sciences at the Microscale and Department of Modern Physics, University of Science and Technology of China, Hefei 230026, China}
\affiliation{
CAS Key Laboratory of Microscale Magnetic Resonance, University of Science and Technology of China, Hefei 230026, China}
\affiliation{
Synergetic Innovation Center of Quantum Information and Quantum Physics, University of Science and Technology of China, Hefei 230026, China}

\author{Yanjun Hou}
\affiliation{
Hefei National Laboratory for Physical Sciences at the Microscale and Department of Modern Physics, University of Science and Technology of China, Hefei 230026, China}
\affiliation{
CAS Key Laboratory of Microscale Magnetic Resonance, University of Science and Technology of China, Hefei 230026, China}
\affiliation{
Synergetic Innovation Center of Quantum Information and Quantum Physics, University of Science and Technology of China, Hefei 230026, China}

\author{Jiangyu Cui}
\email{cuijiangyu@huawei.com}
\affiliation{Central Research Institute, Huawei Technologies, Shenzhen, 518129, China}

\author{Man-Hong Yung}
\email{yung@sustech.edu.cn}
\affiliation{Central Research Institute, Huawei Technologies, Shenzhen, 518129, China}
\affiliation{
Department of Physics, 
Southern University of Science and Technology, Shenzhen 518055, China.}
\affiliation{
Shenzhen Institute for Quantum Science and Engineering, 
Southern University of Science and Technology, Shenzhen 518055, China.}
\affiliation{
Guangdong Provincial Key Laboratory of Quantum Science and Engineering,
Southern University of Science and Technology, Shenzhen 518055, China.}
\affiliation{
Shenzhen Key Laboratory of Quantum Science and Engineering,
Southern University of Science and Technology, Shenzhen, 518055, China.}

\author{Xinhua Peng}
\email{xhpeng@ustc.edu.cn}
\affiliation{
Hefei National Laboratory for Physical Sciences at the Microscale and Department of Modern Physics, University of Science and Technology of China, Hefei 230026, China}
\affiliation{
CAS Key Laboratory of Microscale Magnetic Resonance, University of Science and Technology of China, Hefei 230026, China}
\affiliation{
Synergetic Innovation Center of Quantum Information and Quantum Physics, University of Science and Technology of China, Hefei 230026, China}

\date{\today}

\begin{abstract}

Gradient ascent pulse engineering algorithm (GRAPE) is a typical method to solve quantum optimal control problems.
However, it suffers from an exponential resource in computing the time evolution of quantum systems with the increasing number of qubits, which is a barrier for its application in large-qubit systems.
To mitigate this issue, we propose an iterative GRAPE algorithm (iGRAPE) for preparing a desired quantum state, where the large-scale, resource-consuming optimization problem is decomposed into a set of lower-dimensional optimization subproblems by disentanglement operations.
Consequently these subproblems can be solved in parallel with less computing resources.
For physical platforms such as nuclear magnetic resonance (NMR) and superconducting quantum systems, we show that iGRAPE can provide up to 13-fold speedup over GRAPE when preparing desired quantum states in systems within 12 qubits.
Using a four-qubit NMR system, we also experimentally verify the feasibility of the iGRAPE algorithm.

\end{abstract}

\maketitle

\section*{INTRODUCTION}

In the past few decades, the quantum optimal control (QOC) \cite{krotov1995global,bryson2018applied,werschnik2007quantum,glaser2015training,lloyd2014information} theory has been developed well and stimulates lots of interests in the field of the quantum technology.
Specifically, this theory focuses on the topic for optimal implementation of a target quantum state or desired quantum operation in quantum simulation and quantum sensing as well as scalable quantum computation device \cite{machnes2015gradient,liebermann2016optimal,egger2014adaptive,dolde2014high,platzer2010optimal,goerz2015optimizing,watts2015optimizing}.
To realise this topic in practice, an efficient optimization algorithm is essential.
So far, various numerical optimization algorithms have been developed, including the gradient-based methods such as GRAPE \cite{khaneja2005optimal,larocca2021krylov}, Stochastic Gradient Descent \cite{SGD,turinici2019stochastic}, Krotov algorithm \cite{krotov,goerz2019krotov}, reinforcement learning and their variants \cite{RLxiaoming,niu2019universal,zhang2018automatic,albarran2018measurement,august2018taking,bukov2018reinforcement,chen2019manipulation}, as well as the non gradient-based methods such as Chopped Random Basis \cite{doria2011optimal,rach2015dressing} and Nelder-Mead approach \cite{nelder_mead}.

GRAPE algorithm has attracted much attention among those algorithms.
As it utilizes a direct analytical expression for the gradient, GRAPE can efficiently find a suitable solution in the parameter space with fast convergence speed.
Moreover, in combination with the advanced optimizers (such as BFGS, AdaGrad, and Adam), as well as automatic differentiation technique implemented on GPU \cite{leung2017speedup}, a lot of variants of GRAPE were proposed to improve its performance \cite{ge2021risk,ge2020robust}.
The GRAPE algorithm was originally developed in NMR systems \cite{khaneja2005optimal}, which has now been widely applied to many other quantum platforms \cite{yang2020assessing,dalgaard2020hessian,saywell2020optimal,boutin2017resonator,kwon2021gate}, such as superconducting quantum  circuits\cite{kwon2021gate}, circuit QED\cite{dalgaard2020hessian}, trapped ions\cite{katz2022programmable}, nitrogen-vacancy (NV) centres\cite{rong2015experimental}.
However, this technique relies on dynamic simulations of the quantum systems on the classical computers, which essentially limits its application on large quantum systems.

To mitigate the problem above, here we propose a new variant, named as Iterative Gradient Ascent Pulse Engineering (iGRAPE) algorithm, for finding an optimal solution to the problem of controlling a quantum system from a defined initial state to a desired target state, i.e., a \textit{state-to-state} problem.
Different from the original GRAPE that tackles the dynamics of the entire system, the iGRAPE algorithm adopts the inverse evolution of the target state to the initial state, and gradually decomposes the optimization problem into a set of parallelizable low-dimensional components by using the disentanglement partition systems. This greatly decreases the computing resource for the optimization.  
We apply the algorithm to two typical physical platforms (i.e., NMR and superconducting quantum platforms) to prepare the desired states. The results show that iGRAPE can be up to 13 times faster than GRAPE method for systems within 12 qubits.
In addition, we experimentally prepare the Greenberger–Horne–Zeilinger (GHZ) state using the iGRAPE method on a 4-qubit NMR platform with an experimental fidelity of 98.25\%. 
By the comparison, we found that the iGRAPE algorithm is superior to the GRAPE algorithm in the application of large quantum systems, providing a more possible scheme for further applications of QOC in noisy intermediate scale quantum systems.

\section*{RESULTS}

\subsection*{The iGRAPE algorithm}

A quantum system generally can be described by the Hamiltonian,
\begin{align}
    H(t) = H_{\mathrm{s}}(t) + H_{\mathrm{c}}(t),
\end{align}
where the system Hamiltonian is:
\begin{align}
    H_{\mathrm{s}}(t) = H_{\mathrm{L}} + \sum_{i,j} g_{ij}(t) H_{ij},
\end{align}
and the control-field Hamiltonian reads:
\begin{equation}
    H_{\mathrm{c}}(t) = \sum_{\alpha} u_{\alpha}(t) H_{\alpha}.
    \label{equ:H}
\end{equation}
Here $H_{ij}$ denotes the coupling between qubits $i$ and $j$ with the coupling strength $g_{ij}(t)$ and $H_{\mathrm{L}}$ represents the local terms of the system Hamiltonian; $H_{\alpha}$ represents the control-field Hamiltonian with the amplitude $u_{\alpha}(t)$.

\begin{figure}[htbp]
    \centering
    \includegraphics[width=1\linewidth]{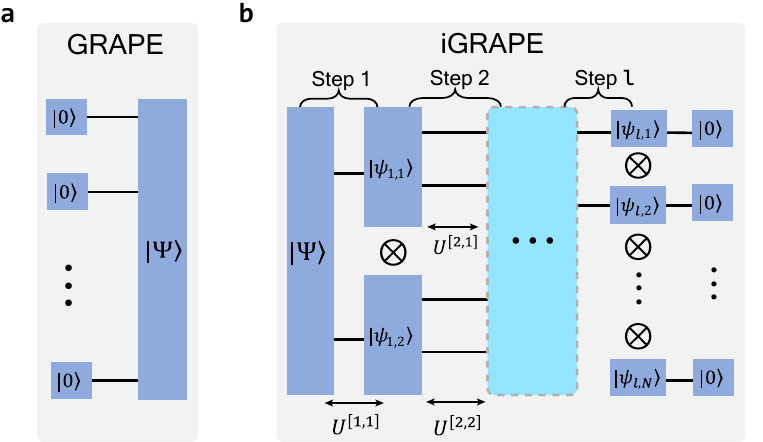}
    \caption{
    {\bfseries Scheme of GRAPE and iGRAPE for state preparation problems when the initial state is $\ket{\mathbb{0}}$.}
    {\bfseries a.} The GRAPE algorithm designs the control field from the initial state $\ket{\mathbb{0}}$ to the target state $\vert \Psi \rangle$.
    {\bfseries b.} The iGRAPE algorithm calculates the reverse-evolution and design the control field from the target state $\ket{\Psi}$ to the initial state $\ket{\mathbb{0}}$.
    }
    \label{fig:1}
\end{figure}

The goal of GRAPE for the state preparation is to design a suitable set of $u_{\alpha}(t)$ for the control pulse to transfer an initial state $\ket{\psi(0)}$, e.g., $\ket{\psi(0)} = \ket{0 \cdots 0} = \ket{\mathbb{0}}$, to a given target state $\ket{\Psi}$ in a specified time $T$.
We linearly discretize the whole time evolution into $K$ segments, i.e., $\Delta t = T/K$.
For the $k$-th segment, the induced temporal-evolution propagator $\mathcal{U}_{k}$ is
\begin{align}
    \mathcal{U}_{k}(\Delta t) = \exp \left\{ -\mathrm{i} \Delta t \left( H_{\mathrm{s}} + \sum_{\alpha} u_{\alpha}(k) H_{\alpha} \right) \right\}.
    \label{Uk}
\end{align}
Here, we assume the system Hamiltonian $H_{\mathrm{s}}$ and the control parameter within the $k$-th segment $u_{\alpha}(k)$ are time-independent.
Consequently, the unitary evolution of the GRAPE algorithm can be written as $U = \Pi_{k=1}^{K} \mathcal{U}_{k}(\Delta t)$, which is optimized to transfer $\ket{\psi(0)}$ to $\ket{\Psi}$.

The key idea of the iGRAPE algorithm is to utilize the reverse evolution transferring the target state $\ket{\Psi}$ to the initial state $\ket{\psi(0)}$, and gradually disentangle $\ket{\Psi}$ into the product of states in subsystems until the complete product state $\ket{\psi(0)}$ is obtained.
As shown in Fig.~\ref{fig:1}b, the operator $U^{[1,1]}$ optimized in algorithm \textit{Step} 1 transfers the target state $\ket{\Psi}$ to the product of the states of its two subsystems $\ket{\psi_{1,1}}$ and $\ket{\psi_{1,2}}$; then in \textit{Step} 2, the optimized operator $U^{[2,1]}$ further transfers $\ket{\psi_{1,1}}$ to the product of the states of its two subsystems $\ket{\psi_{2,1}} \otimes \ket{\psi_{2,2}}$, and likely $U^{[2,2]}$ transfers $\ket{\psi_{1,2}}$ to $\ket{\psi_{2,3}} \otimes \ket{\psi_{2,4}}$; until in \textit{Step} $l$, all the subsystem states become single-qubit states and can be finally transferred to $\ket{\mathbb{0}}$ by applying single-qubit rotations.
Each subsystem state $\ket{\psi_{n,m}}$ can be an arbitrary state in the corresponding Hilbert space.
Let $n$ labels the step index, and $m$ labels the subsystem index in the \textit{Step}.
The whole propagator $U$ can be written as
\begin{align}
    U = \left( \bigotimes_{m} U^{[l+1,m]} \right) \cdots \left( \bigotimes_{m=1}^{2} U^{[2,m]} \right) \cdot U^{[1,1]},
\end{align}
where the target $U \ket{\Psi} = \ket{\mathbb{0}}$, thus $ \ket{\Psi} = U^{\dagger} \ket{\mathbb{0}}$.

The numerical optimization is applied to each step for engineering the pulse sequence to implement the operation $U^{[n,m]}$.
Like that in GRAPE, we linearly discretize the whole time evolution of $U^{[n,m]}$ into $K_{n}$ segments.
The $k$-th temporal-evolution propagator $\mathcal{U}_{k}^{[n,m]}$ is:
\begin{align}
    \mathcal{U}_{k}^{[n,m]} = \exp \left\{ \mathrm{i} \Delta t \left( H_{\mathrm{s}} + \sum_{\alpha} u_{\alpha}^{[n,m]}(k) H_{\alpha} \right) \right\}.
    \label{equ:evolve}
 \end{align}
Hence $U^{[n,m]} = \Pi^{K_{n}}_{k=1} \mathcal{U}_{k}^{[n,m]}$ and $u_{\alpha}^{[n,m]}(k)$ are the optimized control parameters.
Note that $\mathcal{U}_{k}^{[n,m]}$ is different from Eq.~\eqref{Uk}, here we set $- \Delta t \to \Delta t$ to guarantee the physical implementation of $U^{\dagger}$ in the final pulse sequence.
Consequently, $U^{[n,m]\dagger}$ is directly realized by reversing the pulse sequence engineered for $U^{[n,m]}$. 
Via the operation $U^{[n,m]}$, the system is disentangled into two subsystems: 
\begin{align}
    U^{[n,m]}\ket{\psi_{n-1,m}} = \ket{\psi_{n,2m-1}} \otimes \ket{\psi_{n,2m}}.
    \label{equ:split}
\end{align}
In order to further drive each subsystem independently in the following steps, one needs to turn off all the couplings between these subsystems at the end of each \textit{Step}.
By labeling the two subsystems in Eq.~(\ref{equ:split}) as subsystem A and subsystem B, and setting the cost function as $L_{t} = 1-\tr(\rho^{A} \rho^{A})$, with $\rho^{A} = \tr_{\mathrm{B}}(U^{[n,m]}\ket{\psi_{n-1,m}} \bra{\psi_{n-1,m}}U^{[n,m]\dagger})$.
One can obtain a pure state $\rho^{A}$ by minimizing $L_{t}$.
The optimized parameters $u_{\alpha}(k)$ are updated through the gradient descent rule $u_{\alpha}(k) \gets u_{\alpha}(k) - \omega \cdot \partial L_{t} / \partial u_{\alpha}(k)$ (see Methods).

The iGRAPE scheme also relies on the partitioning of the system in each \textit{Step}, which can be suitably chosen according to the target state and the system Hamiltonian.
In the following, taking GHZ-state preparation as the task, we will benchmark the two algorithms on two physical systems: the superconducting and the NMR quantum-computing systems.

\subsection*{Benchmark on superconducting quantum systems with tunable couplings}

\begin{figure}
    \centering
    \includegraphics[width=1\linewidth]{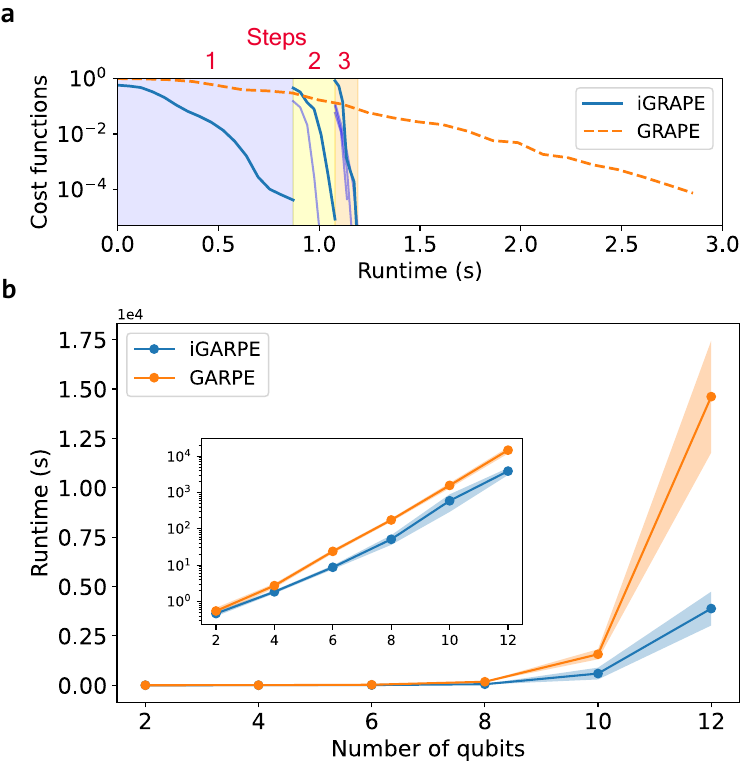}
    \caption{
    {\bfseries Benchmark of iGRAPE and GRAPE algorithms for GHZ-state preparation on superconducting quantum systems.}
    {\bfseries a.} Varying of cost functions during the optimization process of iGRAPE and GRAPE algorithms on a 4-qubit case.
    {\bfseries b.} Runtime vs. Number of qubits. Each point is an average over 20 random selections of a set of the initial control parameters for the different qubit number (see SI). The colored band-region accounts for the statistical distribution. The inset is a semi-log plot.
    }
    \label{fig:superconduct}
\end{figure}

Our benchmark of the iGRAPE algorithm is first performed on a 1D 12-superconducting-qubit chain, where the system Hamiltonian $H_{\text{SC}}$ and the corresponding control field Hamiltonian $H_{\mathrm{c}}$ can be described as \cite{gong2019genuine}:
\begin{equation}
\begin{split}
 H_{\text{SC}} &= \sum_{j} \left( \omega_j \hat{n}_j + \frac{\eta_j}{2} \hat{n}_{j} \left( \hat{n}_{j} -1 \right) \right) \\
 &\quad + \sum_{j} \left( g_{j}(t) \left( \hat{a}^{\dagger}_{j} \hat{a}_{j+1} + \hat{a}_{j} \hat{a}^{\dagger}_{j+1} \right) \right), \\
 H_{\mathrm{c}}(t) &= \sum_{j} \left( u_{xj}(t) \left( \hat{a}_{j} + \hat{a}^{\dagger}_{j} \right) + u_{yj}(t) \mathrm{i} \left( \hat{a}_{j} -\hat{a}^{\dagger}_{j} \right) \right),
\end{split}
\end{equation}
where $\hat{n}$ is the number operator, $\hat{a}_{j}^{\dagger}$ ($\hat{a}_{j}$) is the creation (annihilation) operator, $\omega_{j}$ and $\eta_{j}$ are the transition frequency and the anharmonicity of the $j$-th qubit, respectively, $g_j(t)$ denote the interaction strength between $j$-th and $(j+1)$-th qubits.
Each qubit can be full controlled by individual capacitively coupled microwave control lines (XY), and $u_{xj}, u_{yj}$ are the amplitudes of the control fields.

Fig.~\ref{fig:superconduct}a shows the varying of cost functions during the optimization process of iGRAPE and GRAPE algorithms on a 4-qubit case for a GHZ-state preparation, which consists of three \textit{Steps}:
\begin{align}
\begin{split}
    &\text{Step 1: } \ket{\text{GHZ}} \xrightarrow[]{U^{[1,1]}} \ket{\psi_{1,1}} \otimes \ket{\psi_{1,2}},\\
    &\text{Step 2: } \begin{cases}
    \ket{\psi_{1,1}} &\xrightarrow{U^{[2,1]}} \ket{\psi_{2,1}} \otimes \ket{\psi_{2,2}}, \\
    \ket{\psi_{1,2}} &\xrightarrow{U^{[2,2]}} \ket{\psi_{2,3}}  \ket{\psi_{2,4}},
    \end{cases}\\
    &\text{Step 3: } \ket{\psi_{2,i}} \xrightarrow{U^{[3,i]}} \ket{0},~ i \in [1,4].
\end{split}
\end{align}
Here $\ket{\psi_{1,1}}$ and $\ket{\psi_{1,2}}$ are 2-qubit states, and $\ket{\psi_{2,i}} (i =1,\cdots,4)$ are single-qubit states.
$U^{[1,1]}$, $U^{[2,i]} (i=1,2)$ and $U^{[3,i]}, (i=1,\cdots,4)$ are, respectively, $16 \times 16$, $4 \times 4 $ and $2 \times 2 $ unitary operations.
Although the cost function for each training curve is different (see eq.(\ref{equ:origin_cost}) in Methods), they are all in the range of $[0,1]$.
It can bee seen that the iGRAPE algorithm is fast converged in each \textit{Step} (denoted by blue solid lines), while the GRAPE algorithm takes a longer time to the final convergence (denoted by the red dashed-line).
For \textit{Step} 2 and \textit{Step} 3, the optimization tasks in each \textit{Step} can be simultaneously optimized in lower-dimensional Hilbert spaces.
Note that on superconducting systems, $g_j(t)$ can be controlled off and on, thus in \textit{Steps} 2 and 3, the couplings between the different subsystems are controlled off. 

Using the L-BFGS-B optimization algorithm \cite{l-bfgs-b}, Fig.~\ref{fig:superconduct}b shows that the relationship between the running time of the iGRAPE algorithm and the qubit number on the 1D 12-qubit superconducting-qubit chain, in contrast to the case of the GRAPE algorithm.
It can be seen from Fig.~\ref{fig:superconduct}b that the time consumption of iGRAPE (in blue dots) are less than those of GRAPE (in red dots), and the superiority is more significant as the qubit number grows.
We observe a 5-fold iGRAPE speedup for 12-qubit case.
We also show the semi-log plot in the inset of Fig.~\ref{fig:superconduct}b.
Although the runtime of iGRAPE still grows exponentially with the system size, the exponential index is reduced to 0.398 from 0.446 in the case of GRAPE.

\subsection*{Benchmark on NMR quantum systems with always-on couplings}

\begin{figure*}
    \centering
    \includegraphics[width=1\linewidth]{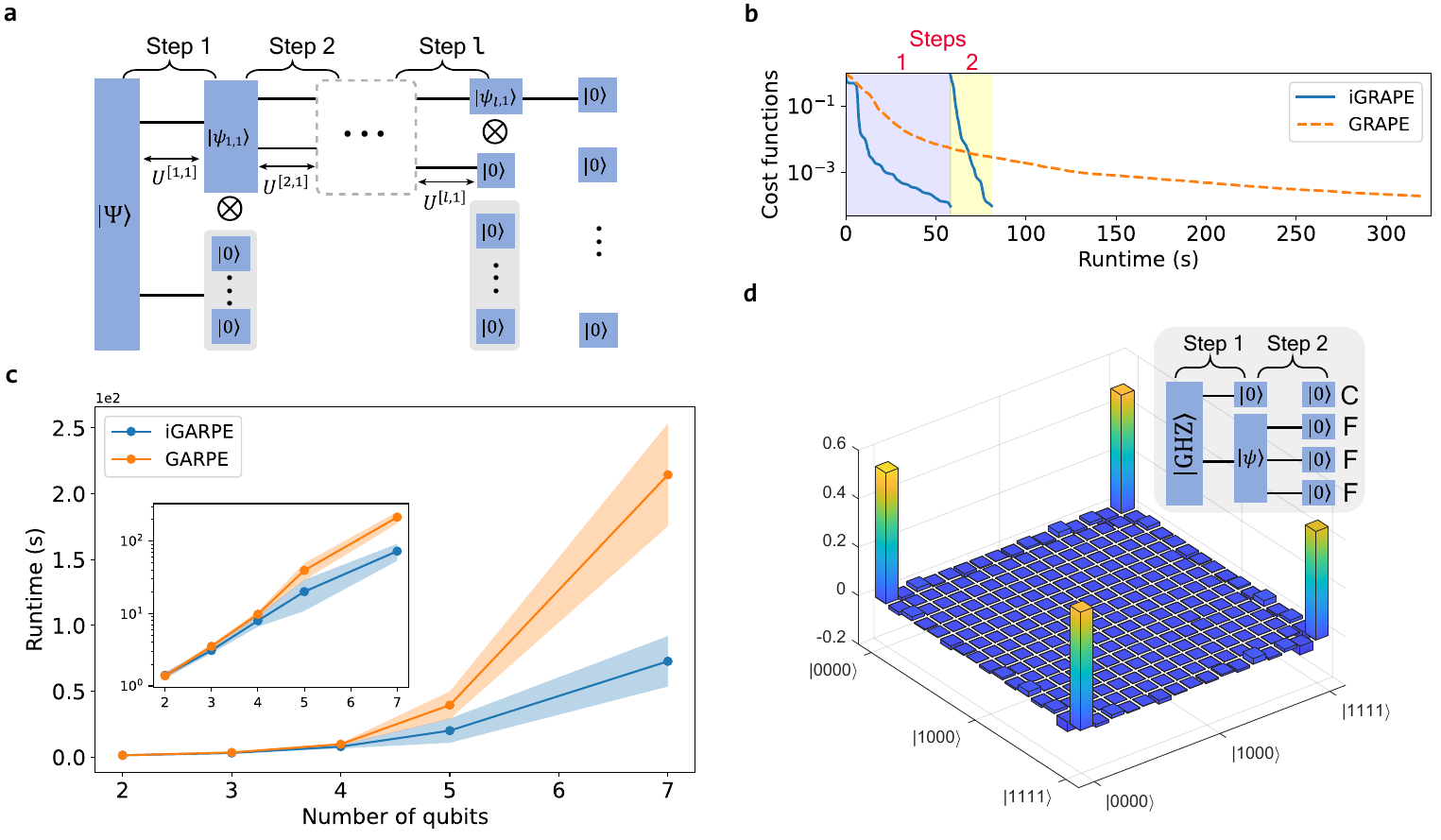}
    \caption{
    {\bfseries Benchmark of iGRAPE and GRAPE algorithms for GHZ-state preparation on NMR quantum systems.}
    {\bfseries a.} Revised scheme of iGRAPE for the coupling always-on NMR systems.
    {\bfseries b.} Evolution of cost functions during the running of iGRAPE and GRAPE algorithms for the 7-qubit NMR quantum system.
    {\bfseries c.} Runtime vs. Number of qubits. Each point is an average over 20 random selections of a set of the initial control parameters on the NMR systems with the different qubit number (see SI). The colored band-region accounts for the statistical distribution. The inset is a semi-log plot.
    {\bfseries d.} Experimentally reconstructed density matrix (real part) for the prepared GHZ state by the iGRAPE algorithm on the 4-qubit NMR system ($^{13}$C-iodotriuroethylene, see SI). The image part is almost zero ($<0.06$).
    }
    \label{fig:nmr}
\end{figure*}

For NMR quantum systems, the system Hamiltonian and the control-field Hamiltonian can be, respectively, written as
\begin{align}
\begin{split}
    H_{\text{NMR}} &= \sum_{j} \left( \pi \nu_{j} \hat{\sigma}_{z}^{(j)} \right) + \sum_{i<j} \left( \frac{\pi}{2} J_{ij} \hat{\sigma}_{z}^{(i)} \hat{\sigma}_{z}^{(j)} \right), \\
    H_{\mathrm{c}}(t) &= \sum_{j} \left( \pi u_{xj}(t) \hat{\sigma}_{x}^{(j)} + \pi u_{yj}(t) \hat{\sigma}_{y}^{(j)} \right).
    \label{equ:HNMR}
\end{split}
\end{align}
Here $H_{\text{NMR}}$ is time-independent, $\nu_{j}$ represents the chemical shift of the $j$-th spin, $J_{ij}$ is the scalar coupling strength between two spins, and $u_{xj}$ and $u_{yj}$ denote, respectively, the control radio-frequency (rf) fields along $x$ and $y$ direction.
In the Supplementary Information, we show the parameters $\nu_{i}$ and $J_{ij}$ for some NMR samples.
Different from the superconducting quantum systems mentioned above, the coupling terms $J_{ij}$ are always on, which should be taken consideration into the optimizations in each \textit{Step}.
In order to avoid the evolution under the couplings between two subspaces in the optimization of the following \textit{Steps}, we set $\ket{\psi_{n,2m}}$ in Eq.~(\ref{equ:split}) as $\ket{\mathbb{0}}$, as shown in Fig.~\ref{fig:nmr}a,
\begin{align}
    U^{[n,m]}\ket{\psi_{n-1,m}} = \ket{\psi_{n,2m-1}} \otimes \ket{\mathbb{0}}.
\end{align}
In the following \textit{Step}, the control fields to be optimized are only performed on the $\ket{\psi_{n,2m-1}}$ state.
For instance, if we consider the NMR sample Crotonic acid which can be regarded as a 7-qubit quantum simulator \cite{knill2000algorithmic}, we divide the system into two subsystems: four $^{13}$C spins and three $^{1}$H spins.
The varying of cost functions in the optimization processes for a GHZ-state preparation is shown in Fig.~\ref{fig:nmr}b, and the algorithm is designed as:
\begin{align}
\begin{split}
    &\text{Step 1: } \ket{\text{GHZ}} \xrightarrow[]{U^{[1,1]}} \ket{\psi_{1,1}}_{\mathrm{C}} \ket{\mathbb{0}}_{\text{H}},\\
    &\text{Step 2: } \ket{\psi_{1,1}}_{\mathrm{C}} \xrightarrow{U^{[2,1]}} \ket{\mathbb{0}}_{\mathrm{C}}.
\end{split}
\end{align}
A similar result was observed as the case in Fig.~\ref{fig:superconduct}a.

Likely in the case of superconducting systems, we inspect how the runtime of the iGRAPE algorithm scales with the number of qubits for GHZ-state preparation.
The benchmark results are shown in Fig.~\ref{fig:nmr}c.
In general, the iGRAPE algorithm (in blue dots) has better performance than the traditional GRAPE algorithm (in red dots).
Significantly, the advantages of iGRAPE are more obvious in larger quantum systems, e.g. a 4-fold improvement is achieved over GRAPE in the 7-qubit case.
The inset shows the exponential index in iGRAPE is reduced to 0.349 from 0.449 in GRAPE.

\subsection*{Experimental verification}

To verify the feasibility of the iGRAPE algorithm in the experiments, we employed $^{13}$C-iodotriuroethylene dissolved in d-chloroform as a 4-qubit quantum simulator, consisting of one $^{13}$C and three $^{19}$F nuclear spins (see SI for the details of this sample) \cite{li2017measuring, chen2021experimental}.
Experiments were performed on a Bruker Avance III 400 MHz spectrometer at room temperature.
The system, initially at the thermal equilibrium state, is first prepared to a pseudo-pure state (PPS) $\hat{\rho}_{\mathrm{pps}} = [(1-\epsilon) / 16] \mathbb{1} + \epsilon \ketbra{\mathbb{0}}$ by using the selective-transition approach \cite{peng2001preparation} with the polarization $\epsilon \approx 10^{-5}$.
The experimental fidelity of $\hat{\rho}_{\mathrm{pps}}$ is about $99.29\%$ (see the Supplementary Information for details).
According to the iGRAPE method, the target GHZ state is first transferred into a product state $\ket{0}_{\mathrm{C}} \ket{\psi}_{\mathrm{F}}$ by designing a pulse sequence in \textit{Step} 1.
Here $\ket{\psi}_{\mathrm{F}}$ is a 3-qubit state for three $^{19}$F spins.
In \textit{Step} 2, we search a pulse sequence to transfer three $^{1}$F spins from $\ket{\psi}_{\mathrm{F}}$ into the state $\ket{\mathbb{0}}$.
The diagram of the iGRAPE process for the 4-qubit case is shown in Fig.~\ref{fig:nmr}d.
Consequently, the final pulse sequence to be tested in the experiment is generated by reversing the whole pulse sequence.
After applying the pulse sequence on the PPS, we then tomography the final state prepared in the experiment.
The tomography result is also shown in Fig.~\ref{fig:nmr}d, with the final state (GHZ state) fidelity 98.25\%.

\section*{DISCUSSION}

\begin{figure*}[htbp]
    \centering
    \includegraphics[width=1\linewidth]{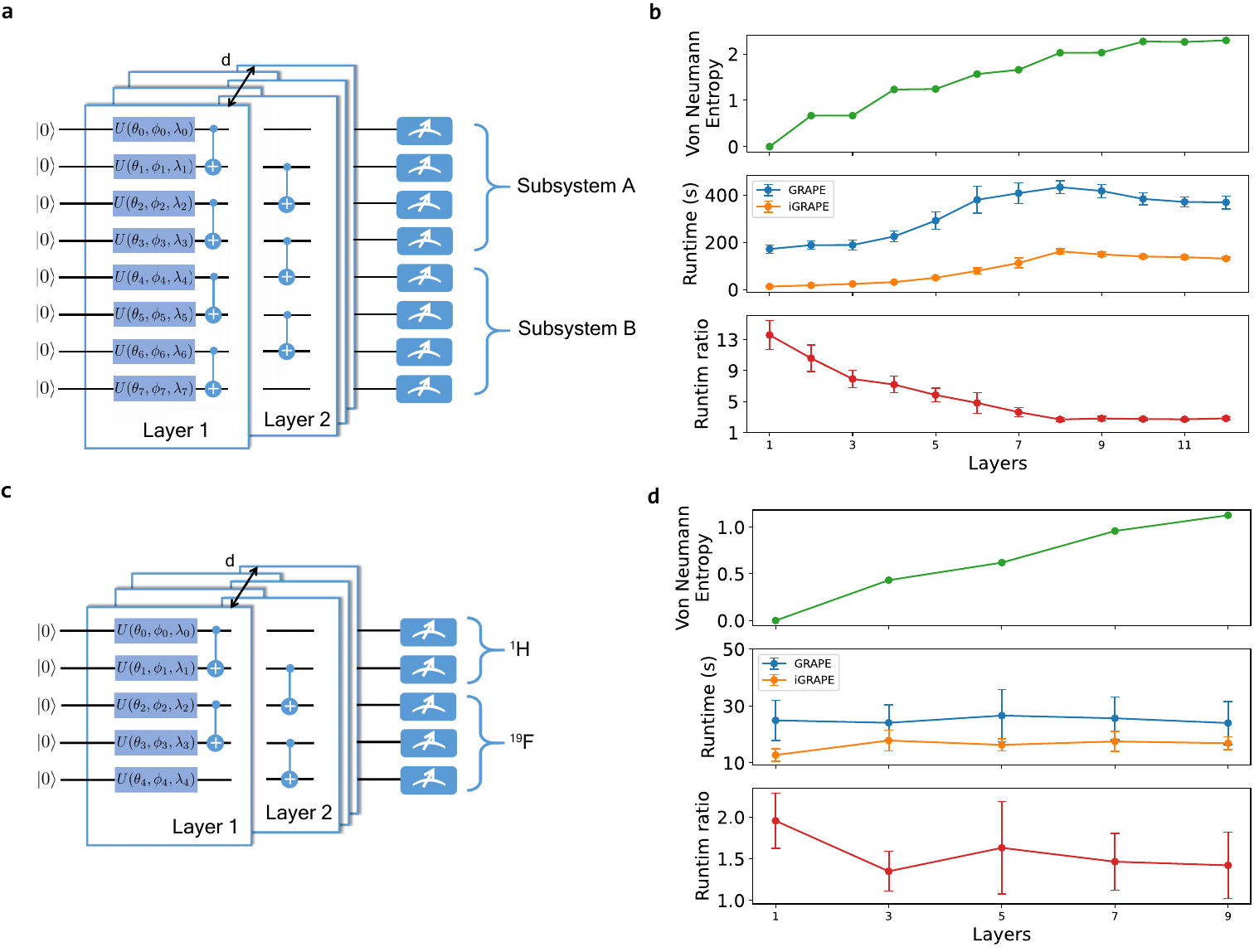}
    \caption{
    {\bfseries Benchmark of iGRAPE and GRAPE algorithms for arbitrary quantum states preparation. Target states are generated by parameterized quantum circuits (PQC).}
    {\bfseries a.} PQC that generates arbitrary 8-qubit quantum states for the superconducting quantum system, where $d$ represents the number of layers.
    {\bfseries b.} The von Neumann entropy, the runtime for both algorithms, and the runtime ratios (GRAPE/iGRAPE) vs. different layers of circuit. Twenty sets of different random parameters have been generated for each circuit, with each point and its error bar representing the mean value and corresponding standard deviation, respectively.
    {\bfseries c.} PQC that generates arbitrary 5-qubit quantum states for the NMR quantum system and
    {\bfseries d.} The corresponding von Neumann entropy, runtime, and the runtime ratios vs. different layers of circuit.
    }
    \label{fig:pqc_test}
\end{figure*}

From the benchmark experiments, we find that the iGRAPE algorithm outperforms the GRAPE algorithm in the state preparation problems, because of the following reasons.
Firstly, the cost functions in the iGRAPE algorithm are only concerned with the disentanglement of the quantum state, regardless of its specific form.
Therefore, the final state is not unique, which makes the optimization process easier to converge.
Secondly, by the disentanglement process, the optimized problem in the iGRAPE algorithm is divided into the problems of the lower-denominational subsystems in each step.
Thirdly, except for the first \textit{Step}, the subsequent optimizations in each \textit{Step} can be carried out in parallel.

In order to further illustrate the validity of the iGRAPE algorithm, we shall discuss its performance for preparing arbitrary quantum states.
We use parameterized quantum circuit (PQC) to generate a set of arbitrary quantum states, which are the target states to prepare.
Since the goal of each \textit{Step} in the iGRAPE algorithm is to transfer the target state into the products of subsystems, the entanglement between the subsystems of the target state might have a significant effect on the runtime of the algorithm, especially in the first \textit{Step}.
Using the von Neumann entropy as a measure of entanglement \cite{ryu2006aspects}, we can benchmark the performance of the iGRAPE algorithm for the target states with different amount of entanglement.

Fig.~\ref{fig:pqc_test}a and Fig.~\ref{fig:pqc_test}b show the PQC for generating 8-qubit arbitrary state and the corresponding algorithm performance for the superconducting quantum system.
Fig.~\ref{fig:pqc_test}c, d show the PQC to generate 5-qubit arbitrary quantum states and the algorithm performance for the NMR quantum platform.
The general $U$ gate in Fig.~\ref{fig:pqc_test} has the form
\begin{align}
    U(\theta, \phi, \lambda) = 
    \begin{pmatrix}
        \cos(\frac{\theta}{2}) & -\mathrm{e}^{\mathrm{i}\lambda} \sin(\frac{\theta}{2}) \\
        -\mathrm{e}^{\mathrm{i}\phi} \sin(\frac{\theta}{2}) & -\mathrm{e}^{\mathrm{i}\lambda + \mathrm{i}\phi} \cos(\frac{\theta}{2})
    \end{pmatrix},
\end{align}
where $\theta \in [0,\pi],~ \phi \in [0,2\pi],~ \lambda \in [0,2\pi]$, and $\{ \theta_{i}, \phi_{i}, \lambda_{i} \}$ are randomly chosen in the range.
An arbitrary pure state $\ket{\psi}$ of a composite system AB can be described by its Schmidt decomposition:
\begin{align}
    \ket{\psi} = \sum_{i} \alpha_{i} \ket{i_{\mathrm{A}}} \ket{i_{\mathrm{B}}},
\end{align}
where $\ket{i_{\mathrm{A}}}$ and $\ket{i_{\mathrm{B}}}$ are orthonormal states for subsystems A and B, respectively.
For pure states, the von Neumann entropy $S$ of the reduced states $\rho_{\mathrm{A}}$ and $\rho_{\mathrm{B}}$ is a well-defined measure of entanglement,
\begin{align}
    S(\rho_{\mathrm{A}}) = S(\rho_\mathrm{B}) = -\sum_{i} |\alpha_{i}|^{2} \log \left( |\alpha_{i}|^{2} \right),
\end{align}
and this is zero if and only if $\ket{\psi}$ is a product state (not entangled).
In Fig.~\ref{fig:pqc_test} b, we observe a 2.7$\sim$13.6-fold speedup of iGRAPE for 8-qubit arbitrary quantum states of superconducting quantum system.
For the 1-bromo-2,4,5-trifluorobenz sample, we divide the 5-spin system (shown in Fig.~\ref{fig:pqc_test}c) into two $^{1}$H spins as the subsystem A and three $^{19}$F spins as the subsystem B.
From Fig.~\ref{fig:pqc_test} d, we observe that iGRAPE provides 1.5$\sim$2-fold speedup for 5-qubit arbitrary quantum states of NMR quantum system.
As the number of layers increases, the von Neumann entropy of subsystems A and B in the final state goes up.
We find that the two algorithms have similar fluctuation trends for the quantum states obtained under different circuit layers.
The runtime advantage of iGRAPE algorithm tends to be stable with the increase in circuit depth.

If the coupling terms $g_{ij}$ in system Hamiltonian are fixed and non-zero, in ZZ coupling case, we can set one of the subsystems into $\ket{\mathbb{0}}$ to let the other subsystem evolves independently (see Methods).
If the coupled Hamiltonian $H_{ij}$ has other forms \cite{Nielsen}, we can use a similar idea to transform one of the subsystems into the eigenstate of $H_{ij}$ such that operations on the other subsystems only give that subsystem a global phase factor.
At present, for pulse sequence with $K$ unitary operators, the number of operators $K_{n}$ assigned to the $n$-th \textit{Step} in iGRAPE algorithm depend mainly on the system and control Hamiltonian.
If the entanglement information of the target state is taken into account, we may be able to design a better scheme to reduce the number of \textit{Steps}.
For other state transfer problems, such as transferring an arbitrary state $\ket{\psi}$ to another arbitrary state $\ket{\psi'}$, we can set state $\ket{\mathbb{0}}$ as an intermediate state, then use iGRAPE to generate a pulse sequence from $\ket{\psi}$ to $\ket{\mathbb{0}}$, and use it again to generate another pulse sequence from $\ket{\mathbb{0}}$ to $\ket{\psi'}$.
Currently, the iGRAPE algorithm is limited to state transfer problems.
In the future, we will try to enlarge the range of iGRAPE algorithm, make its possible application in the broader field of quantum optimal control, such as unitary operation preparation, etc.

In conclusion, we propose a novel quantum optimal control scheme for the state transfer problems and benchmark the algorithm on different sizes of GHZ state preparations, as well as experimental validation on a 4-qubit NMR quantum processor.
We compare the iGRAPE algorithm with the original GRAPE algorithm from the perspective of preparing quantum states with different entanglement degrees, and the results show that the iGRAPE algorithm has advantages in runtime.
The advantages of the algorithm are very important for practical applications on real quantum systems in the noisy intermediate-scale quantum (NISQ) era.

\section*{METHODS}

\subsection*{Algorithm benchmarking}

For both GRAPE and iGRAPE in this paper, we use the L-BFGS-B optimization algorithm \cite{l-bfgs-b} for pulse optimization and achieve a minimum state fidelity of 99.7\% in all of our examples.
The CPU model we used was an Intel$^\circledR$ Xeon$^\circledR$ CPU E5-2620 V3 @ 2.4GHz.
The NMR sample information corresponding to different system sizes and the experimental parameters for superconducting quantum systems are provided in the Supplementary Information.
The pseudocode for iGRAPE is shown in Algorithm \ref{alg:igrape}.

\begin{figure*}
\begin{minipage}{\linewidth}
\begin{algorithm}[H]
    \caption{The iGRAPE algorithm}
    \label{alg:igrape}
    \begin{algorithmic}[1]
    \Statex \textbf{Input:} Target state $\ket{\Psi}$
    \Statex \qquad\quad ~ Set tolerance $\epsilon_{0}$
    \Statex \textbf{Output:} Pulse sequence $U$ that satisfies $U^{\dagger} \ket{\mathbb{0}} = \ket{\Psi}$
    \State Set $n=1$, $\ket{\psi_{n-1,1}} = \ket{\Psi}$
    \While{$\exists a$ s.t. $\ket{\psi_{n-1,a}} \neq$ single qubit state} \Comment{Outer loop: divide the quantum system}
        \For{$m$ in $\{ m | \ket{\psi_{n-1,m}} \neq \text{single qubit state}\}$} \Comment{Inner loop: generate pulse sequence for each subsystem}
            \State Guess initial controls $u_{\alpha}^{[n,m]}(k)$
            \State Evolution: $\ket{\phi_{n,m}} = \mathcal{U}_{K_{n}}^{[n,m]}...\mathcal{U}_{1}^{[n,m]} \ket{\psi_{n-1,m}}$
            \State Calculate cost function $L$
            \While{$L \geq \epsilon_{0}$} \Comment{Minimize cost function $L$}
                \State Calculate gradient: $\frac{\partial L}{\partial u_{\alpha}^{[n,m]}(k)}$
                \State Update: $u_{\alpha}^{[n,m]}(k) \gets u_{\alpha}^{[n,m]}(k) - \omega\frac{\partial L}{\partial u_{\alpha}^{[n,m]}(k)}$
                \State Evolution: $\ket{\phi_{n,m}} = \mathcal{U}_{K_{n}}^{[n,m]}...\mathcal{U}_{1}^{[n,m]} \ket{\psi_{n-1,m}}$
                \State Recalculate cost function $L$
            \EndWhile
            \State Pulse sequence for the subsystem is $U^{[n,m]} = \Pi^{K_{n}}_{k=1} \mathcal{U}_{k}^{[n,m]}$
        \EndFor
        \State $n \gets n+1$
    \EndWhile
    \For{$m$ in $\{ m | \ket{\psi_{n-1,m}} \neq \ket{\mathbb{0}}\}$}
    \State Generate single qubit rotation $U^{[n,m]}$ s.t. $U^{[n,m]} \ket{\psi_{n-1,m}} = \ket{\mathbb{0}}$
    \EndFor
    \State The whole pulse sequence is $U = \left( \bigotimes_{m} U^{[l+1,m]} \right) \cdots \left( \bigotimes_{m=1}^{2} U^{[2,m]} \right) \cdot U^{[1,1]}$
    \end{algorithmic}
\end{algorithm}
\end{minipage}
\end{figure*}

\subsection*{Cost functions in the iGRAPE algorithm}

From the Algorithm \ref{alg:igrape} step 3 to 13, we generate an unitary operator $U^{[n,m]}$ such that
\begin{align}
    U^{[n,m]}\ket{\psi_{n-1,m}} = \ket{\phi_{n,m}} = \ket{\psi_{n,2m-1}}_{\mathrm{A}} \otimes \ket{\psi_{n,2m}}_{\mathrm{B}},
\end{align}
with $U^{[n,m]} = \mathcal{U}_{K_{n}}^{[n,m]} \cdots \mathcal{U}_{1}^{[n,m]}$, which is dependent on the control parameter set $u_{\alpha}^{[n,m]}(k)$.
Suppose the states $\ket{\psi_{n,2m-1}}_{\mathrm{A}}, \ket{\psi_{n,2m}}_{\mathrm{B}}$ are in subsystem A and B, the couplings between the two subsystems should be turned off in the further steps.
In order to find the specific set $u_{\alpha}^{[n,m]}(k)$, the cost function $L_{t}$ is defined as
\begin{align}
    L_{t} = 1 - \tr( \tr_{\mathrm{B}}^{2} \left( \ketbra{\phi_{n,m}} \right) ),
    \label{equ:origin_cost}
\end{align}
minimize $L_{t}$ such that $\rho^{\mathrm{A}} = \tr_{\mathrm{B}} \left( \ket{\phi_{n,m}} \bra{\phi_{n,m}} \right)$ becomes pure state.
The parameters $u_{\alpha}^{[n,m]}(k)$ could be adjusted by the update rule
\begin{align}
    u_{\alpha}^{[n,m]}(k) \gets u_{\alpha}^{[n,m]}(k) - \omega\frac{\partial L_{t}}{\partial u_{\alpha}^{[n,m]}(k)}.
\end{align}
Expand $\rho^{\mathrm{A}}$ such that
\begin{align}
    \rho^{\mathrm{A}} = \sum_{j} \left( \mathbb{1}_{\mathrm{A}} \otimes \bra{j}_{\mathrm{B}} \right) \ket{\phi_{n,m}} \bra{\phi_{n,m}} \left( \mathbb{1}_{\mathrm{A}} \otimes \ket{j}_{\mathrm{B}} \right),
\end{align}
the basis of subsystems A and B are denoted as $\ket{i}_{\mathrm{A}}$ and $\ket{j}_{\mathrm{B}}$.
The gradient can be written as
\begin{align}
    &\frac{\partial L_{t}}{\partial u_{\alpha}^{[n,m]}(k)} = \\
    &\quad 4\Re(\bra{\lambda_{n,m}} \mathrm{i} \Delta t \mathcal{U}_{K_{n}}^{[n,m]}...\mathcal{U}_{k+1}^{[n,m]} H_{\alpha} \mathcal{U}_{k}^{[n,m]}...\mathcal{U}_{1}^{[n,m]} \ket{\psi_{n-1,m}}),
\end{align}
where the elements in vector $\ket{\lambda_{n,m}}$ is given as
\begin{align}
    (\bra{i}_{\mathrm{A}} \bra{j}_{\mathrm{B}}) \ket{\lambda_{n,m}} &= \sum_{i',j'} \bra{\phi_{n,m}} (\ket{i'}_{\mathrm{A}} \ket{j'}_{\mathrm{B}}) \cdot \notag\\
    &(\bra{i}_{\mathrm{A}} \bra{j'}_{\mathrm{B}}) \ket{\phi_{n,m}} (\bra{i'}_{\mathrm{A}} \bra{j}_{\mathrm{B}}) \ket{\phi_{n,m}}.
\end{align}

\subsection*{iGRAPE for NMR quantum systems}

For NMR quantum systems, the coupling terms can be expressed as $J_{ij}\hat{\sigma}_{z}^{(i)}\hat{\sigma}_{z}^{(j)}$, where $J_{ij}$ are always on.
The evolution between two subsystems can be frozen when one of these two subsystems is in the $\ket{\mathbb{0}}$:
\begin{align}
    \mathrm{e}^{-\mathrm{i} Ht} \ket{\mathbb{0}}_{\mathrm{A}} \ket{\psi}_{\mathrm{B}} = \mathrm{e}^{\mathrm{i}\theta} \ket{\mathbb{0}}_{\mathrm{A}} \mathrm{e}^{-\mathrm{i} H_{\text{B}}t} \ket{\psi}_{\mathrm{B}},
    \label{equ:fixed_coupling}
\end{align}
where $H$ denotes the whole-system Hamiltonian and $H_{\text{B}}$ denotes the Hamiltonian for the subsystem B:
\begin{align}
    \begin{split}
    H_{\text{B}} = &\frac{\pi}{2} \left[ \sum_{i} \left( 2\nu_{i} + \sum_{p<i} J_{pi} \right) \hat{\sigma}_{z}^{(i)} + \sum_{i<j} J_{ij} \hat{\sigma}_{z}^{(i)} \hat{\sigma}_{z}^{(j)} \right] \\
    &+ H_{\text{B, control}},
    \end{split}
    \label{equ:NMR_sub}
\end{align}
and $\mathrm{e}^{\mathrm{i}\theta}$ is a global phase.
$\nu_{i}$ and $J_{ij}$ represent the sample's chemical shift and its scalar coupling strength, respectively.
When the state of subsystem A reaches to $\ket{\mathbb{0}}_{\mathrm{A}}$, the coupling Hamiltonian $J_{ij} \hat{\sigma}_{z}^{(i)} \hat{\sigma}_{z}^{(j)}$ will not change the state of the subsystem A except for a global phase.
In this situation, the cost function $L_{f}$ can be expressed as: 
\begin{align}
    L_{f} = 1 - \tr(\rho_{0} \rho^{\mathrm{A}}),
\end{align}
where $\rho_{0}$ denotes the state $\ketbra{\mathbb{0}}_{\mathrm{A}}$ in subsystem A.
The gradient is
\begin{align}
    &\frac{\partial L_{f}}{\partial u_{\alpha}(k)} = \\
    &\quad 2\Re(\bra{\eta_{n,m}} \mathrm{i} \Delta t \mathcal{U}_{K_{n}}^{[n,m]}...\mathcal{U}_{k+1}^{[n,m]} H_{\alpha} \mathcal{U}_{k}^{[n,m]}...\mathcal{U}_{1}^{[n,m]} \ket{\psi_{n-1,m}}),
\end{align}
where the elements in vector $\ket{\eta_{n,m}}$ is defined as:
\begin{align}
    (\bra{\mathbb{0}}_{\mathrm{A}} \bra{j}_{\mathrm{B}}) \ket{\eta_{n,m}} = (\bra{\mathbb{0}}_{\mathrm{A}} \bra{j}_{\mathrm{B}}) \ket{\phi_{n,m}}.
\end{align}

The iGRAPE algorithm process for the NMR experiment shown in Fig.~\ref{fig:nmr}d contains two algorithm \textit{Steps}:
\begin{align}
\begin{split}
    &\text{Step 1: } \ket{\text{GHZ}} \xrightarrow[]{U^{[1]}} \ket{0}_{\mathrm{C}} \ket{\psi}_{\text{F}},\\
    &\text{Step 2: } \ket{\psi}_{\text{F}} \xrightarrow{U^{[2]}} \ket{\mathbb{0}}_{\text{F}}.
\end{split}
\end{align}
The fidelity of a pure state $\ket{\psi}$ and a density matrix $\rho$ is defined as:
\begin{align}
    F(\ket{\psi},\rho) = \bra{\psi}\rho\ket{\psi},
\end{align}
which is used to calculate the final state fidelity of the experiment.

\section*{DATA AVAILABILITY}
Data generated and analyzed during the current study are available from the corresponding author upon reasonable request.

\section*{CODE AVAILABILITY}
Source codes used to generate the plots are available from the corresponding author upon request.

\section*{ACKNOWLEDGEMENTS}
This work is supported by the National Key R \& D Program of China (Grants No. 2018YFA0306600 and 2016YFA0301203), the National Science Foundation of China (Grants No. 11822502, 11974125 and 11927811), Anhui Initiative in Quantum Information Technologies (Grant No. AHY050000).
M.H.Y is supported by the National Natural Science Foundation of China (11875160), the Guangdong Innovative and Entrepreneurial Research Team Program (2016ZT06D348), Natural Science Foundation of Guangdong Province (2017B030308003), and Science, Technology and Innovation Commission of Shenzhen Municipality (ZDSYS20170303165926217, JCYJ20170412152620376, JCYJ20170817105046702).

\section*{AUTHOR CONTRIBUTIONS}
X. P. and M. H. Y. conceived the project.
Y. H., M. H. Y. and Y. C. formulated the theoretical framework.
Y. C. designed the experiment.
Y. C., Z. W. and R. L. performed the measurements and analyzed the data.
Z. W., R. L. and X. P. assisted with the experiment.
X. P. supervised the experiment.
Y. C., Y. H., B. Y. W. and Z. W. wrote the manuscript.
All authors contributed to analyzing the data, discussing the results and commented on the writing.

\section*{COMPETING INTERESTS}
The authors declare no competing interests.

\end{document}


\widetext

\title{Supplementary Information for ``Iterative Gradient Ascent Pulse Engineering algorithm for quantum optimal control''}

\author{Yuquan Chen}
\thanks{These authors contribute equally}
\affiliation{
Hefei National Laboratory for Physical Sciences at the Microscale and Department of Modern Physics, University of Science and Technology of China, Hefei 230026, China}
\affiliation{
CAS Key Laboratory of Microscale Magnetic Resonance, University of Science and Technology of China, Hefei 230026, China}
\affiliation{
Synergetic Innovation Center of Quantum Information and Quantum Physics, University of Science and Technology of China, Hefei 230026, China}

\author{Yajie Hao}
\thanks{These authors contribute equally}
\affiliation{
Department of Physics, 
Southern University of Science and Technology, Shenzhen 518055, China.}
\affiliation{
Shenzhen Institute for Quantum Science and Engineering, 
Southern University of Science and Technology, Shenzhen 518055, China.}

\author{Ze Wu}
\affiliation{
Hefei National Laboratory for Physical Sciences at the Microscale and Department of Modern Physics, University of Science and Technology of China, Hefei 230026, China}
\affiliation{
CAS Key Laboratory of Microscale Magnetic Resonance, University of Science and Technology of China, Hefei 230026, China}
\affiliation{
Synergetic Innovation Center of Quantum Information and Quantum Physics, University of Science and Technology of China, Hefei 230026, China}

\author{Bi-Ying Wang}
\affiliation{Central Research Institute, Huawei Technologies, Shenzhen, 518129, China}

\author{Ran Liu}
\affiliation{
Hefei National Laboratory for Physical Sciences at the Microscale and Department of Modern Physics, University of Science and Technology of China, Hefei 230026, China}
\affiliation{
CAS Key Laboratory of Microscale Magnetic Resonance, University of Science and Technology of China, Hefei 230026, China}
\affiliation{
Synergetic Innovation Center of Quantum Information and Quantum Physics, University of Science and Technology of China, Hefei 230026, China}

\author{Yanjun Hou}
\affiliation{
Hefei National Laboratory for Physical Sciences at the Microscale and Department of Modern Physics, University of Science and Technology of China, Hefei 230026, China}
\affiliation{
CAS Key Laboratory of Microscale Magnetic Resonance, University of Science and Technology of China, Hefei 230026, China}
\affiliation{
Synergetic Innovation Center of Quantum Information and Quantum Physics, University of Science and Technology of China, Hefei 230026, China}

\author{Jiangyu Cui}
\email{cuijiangyu@huawei.com}
\affiliation{Central Research Institute, Huawei Technologies, Shenzhen, 518129, China}

\author{Man-Hong Yung}
\email{yung@sustech.edu.cn}
\affiliation{Central Research Institute, Huawei Technologies, Shenzhen, 518129, China}
\affiliation{
Department of Physics, 
Southern University of Science and Technology, Shenzhen 518055, China.}
\affiliation{
Shenzhen Institute for Quantum Science and Engineering, 
Southern University of Science and Technology, Shenzhen 518055, China.}
\affiliation{
Guangdong Provincial Key Laboratory of Quantum Science and Engineering,
Southern University of Science and Technology, Shenzhen 518055, China.}
\affiliation{
Shenzhen Key Laboratory of Quantum Science and Engineering,
Southern University of Science and Technology, Shenzhen, 518055, China.}

\author{Xinhua Peng}
\email{xhpeng@ustc.edu.cn}
\affiliation{
Hefei National Laboratory for Physical Sciences at the Microscale and Department of Modern Physics, University of Science and Technology of China, Hefei 230026, China}
\affiliation{
CAS Key Laboratory of Microscale Magnetic Resonance, University of Science and Technology of China, Hefei 230026, China}
\affiliation{
Synergetic Innovation Center of Quantum Information and Quantum Physics, University of Science and Technology of China, Hefei 230026, China}

\date{\today}

\maketitle

\section*{The GRAPE algorithm}

A quantum system generally can be described by the Hamiltonian,
\begin{align}
    H(t) = H_{\mathrm{s}}(t) + H_{\mathrm{c}}(t),
\end{align}
which is tunable through the time-dependent control Hamiltonian $H_{\mathrm{c}}(t)$.
The goal of quantum optimal control is to prepare the target state or the target unitary by appropriately adjusting $H_{\mathrm{c}}(t)$.
For example, consider the situation where the system is initially in a given state $\ket{\mathbb{0}} = \ket{0...0}$ and we are interested in a dynamics that prepares the target state $\ket{\Psi}$ through transfer time $T$.
In order to search for controls that accomplish the task, a cost function $L$ must be introduced to quantifying the degree of fulfilment,
\begin{align}
    L = 1 - \left| \bra{\Psi} U(T) \ket{\mathbb{0}} \right|^2,
\end{align}
where $U(T)$ is given by the fact that the latter is the solution of 
\begin{align}
    \begin{split}
        \mathrm{i} \frac{\dd U(t)}{\dd t} &= H(t) U(t), \\
        U(0) &= \mathbb{1},
    \end{split}
\end{align}
evaluated at $t=T$.

For GRAPE algorithm, the control Hamiltonian can be expressed as
\begin{align}
    H_{\mathrm{c}}(t) = \sum_{\alpha} u_{\alpha}(t) H_{\alpha},
\end{align}
where a set of external control fields $u_{\alpha}(t)$ acting on the system via control operators $H_{\alpha}$.
We will assume for simplicity that the chosen transfer time $T$ is discretized in $K$ equal steps of duration $\Delta t = T/K$.
During each step, the control amplitudes $u_{\alpha}$ are constant, i.e. during the $k$-th step, the amplitude $u_{\alpha}$ of the $\alpha$-th control Hamiltonian is given by $u_{\alpha}(k)$.
The time-evolution of the system during a time step $k$ is given by the propagator
\begin{align}
    U_{k} = \exp \left\{ -\mathrm{i} \Delta t \left(H_{\mathrm{s}} + \sum_{\alpha} u_{\alpha}(k) H_{\alpha} \right) \right\},
    \label{equ_up}
\end{align}
and the corresponding cost function $L_{g}$ becomes
\begin{align}
    \begin{split}
        L_{g} &= 1 - \left| \bra{\Psi} \ket{\psi(T)} \right|^2, \\
        \ket{\psi(T)} &= U_{K}...U_{1} \ket{\mathbb{0}}.
    \end{split}
\end{align}
Update $u_{\alpha}(k)$ to minimize the cost function $L_{g}$,
\begin{align}
    u_{\alpha}(k) \gets u_{\alpha}(k) - \omega\frac{\partial L_{g}}{\partial u_{\alpha}(k)}.
\end{align}
Here $\omega$ is the learning rate (a small step size).
This forms the basis of the GRAPE algorithm.
The whole process of GRAPE is shown in Algorithm \ref{alg:grape}.

\begin{figure*}
\begin{minipage}{\linewidth}
\begin{algorithm}[H]
\caption{The GRAPE algorithm}
\label{alg:grape}
\begin{algorithmic}[1]
\Statex \textbf{Input:} Target state $\ket{\Psi}$
\Statex \qquad\quad ~ Set tolerance $\epsilon_{0}$
\Statex \textbf{Output:} Control parameters $u_{\alpha}(k)$
\State Initialization: guess initial controls $u_{\alpha}(k)$
\State Evolution: $\ket{\psi(T)} = U_{K}...U_{1} \ket{\mathbb{0}}$
\State Cost function: $L = 1 - \left|\bra{\Psi} \ket{\psi(T)}\right|^2$
\While{$L \geq \epsilon_{0}$}
    \State Calculate gradient $\frac{\partial L}{\partial u_{\alpha}(k)}$
    \State Update: $u_{\alpha}(k) \gets u_{\alpha}(k) - \omega\frac{\partial L}{\partial u_{\alpha}(k)}$
    \State Evolution: $\ket{\psi(T)} = U_{K}...U_{1} \ket{\mathbb{0}}$
    \State Cost function: $L = 1 - \left|\bra{\Psi} \ket{\psi(T)}\right|^2$
\EndWhile
\end{algorithmic}
\end{algorithm}
\end{minipage}
\end{figure*}

\section*{iGRAPE for NMR quantum systems}

The NMR samples used in the journal text are shown in Table \ref{tab:NMR_samples}.
Here we use the subsystem of 3-qubit NMR sample Diethyl fluoromalo as the 2-qubit system.

\begin{center}
\footnotesize
\LTcapwidth=\textwidth
\begin{longtable}{|c|c|c|c|}
    \caption{NMR samples of different system sizes. The molecular parameters of each sample including $\nu_{i}$ (diagonal) and $J_{ij}$ (off-diagonal).}
    \label{tab:NMR_samples}\\
    \hline
    Sample Name & Molecular Formula & System Size & Parameters (Hz) \\
    \hline
    Diethyl fluoromalonate & FCH(COOC$_{2}$H$_{5}$)$_{2}$ & 2 qubits &
        \begin{tabular}{c|cccc}
        & H & F & $T_1$(s) & $T_2$(s)  \\
        \hline
        H & $400M$ && 2.8 & 1.2 \\
        F & 47.6 & $376M$ & 3.1 & 1.3
        \end{tabular} \\
    \hline
    Diethyl fluoromalonate & FCH(COOC$_{2}$H$_{5}$)$_{2}$ & 3 qubits &
        \begin{tabular}{c|ccccc}
        & C & H & F & $T_1$(s) & $T_2$(s)  \\
        \hline
        C & $100M$ &&& 2.9 & 1.1 \\
        H & 160.7 & $400M$ && 2.8 & 1.2 \\
        F & -194.4 & 47.6 & $376M$ & 3.1 & 1.3
        \end{tabular} \\
    \hline
    Carbon-13-iodotrifluroethylene & C$_{2}$F$_{3}$I & 4 qubits &
        \begin{tabular}{c|cccccc}
        & C & F$_{1}$ & F$_{2}$ & F$_{3}$ & $T_{1}$(s) & $T_{2}$(s) \\
        \hline
        C & 15479.88 &&&& 7.9 & 1.22\\
        F$_{1}$ & -297.71 & -33132.45 &&& 6.8 & 0.66\\
        F$_{2}$ & -275.56 & 64.74 & -42682.97 && 4.4 & 0.63 \\
        F$_{2}$ & 39.17 & 51.50 & -129.08 & -56445.71 & 4.8 & 0.61
        \end{tabular} \\
    \hline
    1-bromo-2,4,5-trifluorobenzene & BrC$_{6}$H$_{2}$F$_{3}$ & 5 qubits &
        \begin{tabular}{c|ccccccc}
        & F$_{1}$ & F$_{2}$ & F$_{3}$ & H$_{1}$ & H$_{2}$ & $T_{1}$(s) & $T_{2}^{*}$(ms) \\
        \hline
        F$_{1}$ & -47708 &&&&& 0.8 & 50\\
        F$_{2}$ & -45.5 & -45257 &&&& 0.6 & 50\\
        F$_{3}$ & 135.8 & 323.5 & -37734 &&& 0.8 & 50\\
        H$_{1}$ & 62.1 & 1468.2 & 1811.2 & 2396 && 1.5 & 110\\
        H$_{2}$ & 1781.1 & 122.9 & 60.9 & -10.1 & 2393 & 1.5 & 110
        \end{tabular} \\
    \hline
    Crotonic acid & C$_{4}$H$_{6}$O$_{2}$ & 7 qubits &
        \begin{tabular}{c|ccccccc}
        & C$_{1}$ & C$_{2}$ & C$_{3}$ & C$_{4}$ & H$_{1}$ & H$_{2}$ & H$_{3}$ \\
        \hline
        C$_{1}$ & 1750.3 \\
        C$_{2}$ & 40.8 & 14930.1 \\
        C$_{3}$ & 1.6 & 69.5 & 12199.9 \\
        C$_{4}$ & 8.47 & 1.4 & 71.04 & 17173.7 \\
        H$_{1}$ & 4.0 & 155.6 & -1.8 & 6.5 & 2785.85 \\
        H$_{2}$ & 6.64 & -0.7 & 162.9 & 3.3 & 15.81 & 2320.25 \\
        H$_{3}$ & 128 & -7.1 & 6.6 & -0.9 & 6.9 & -1.7 & 718.487 \\
        \end{tabular} \\
    \hline
\end{longtable}
\end{center}

The duration $\Delta t$ for each propagator $\mathcal{U}_{k}^{[n,m]}$ in journal text is $5\mu$s, and the parameters of shaped pulses are shown in Table \ref{tab:fig2}.
In the iGRAPE algorithm, we empirically decide how to divide the number of pulses in each subsystem.

\begin{table}[H]
    \centering
    \footnotesize
    \begin{tabular}{|c|c|c|c|c|c|}
    \hline
    Number of qubits & 2 & 3 & 4 & 5 & 7 \\
    \hline
    iGRAPE pulse steps & 500, 100 & 800, 200 & 1500, 260 & 2000, 400 & 2800, 1200 \\
    \hline
    GRAPE pulse steps & 600 & 1000 & 1760 & 2400 & 3000 \\
    \hline
    Transfer time (ms) & 3 & 5 & 8.8 & 12 & 20 \\
    \hline
    \end{tabular}
    \caption{Shaped pulses details of states preparation on NMR quantum systems.}
    \label{tab:fig2}
\end{table}

\begin{figure}[htbp]
    \centering
    \includegraphics[width=1\linewidth]{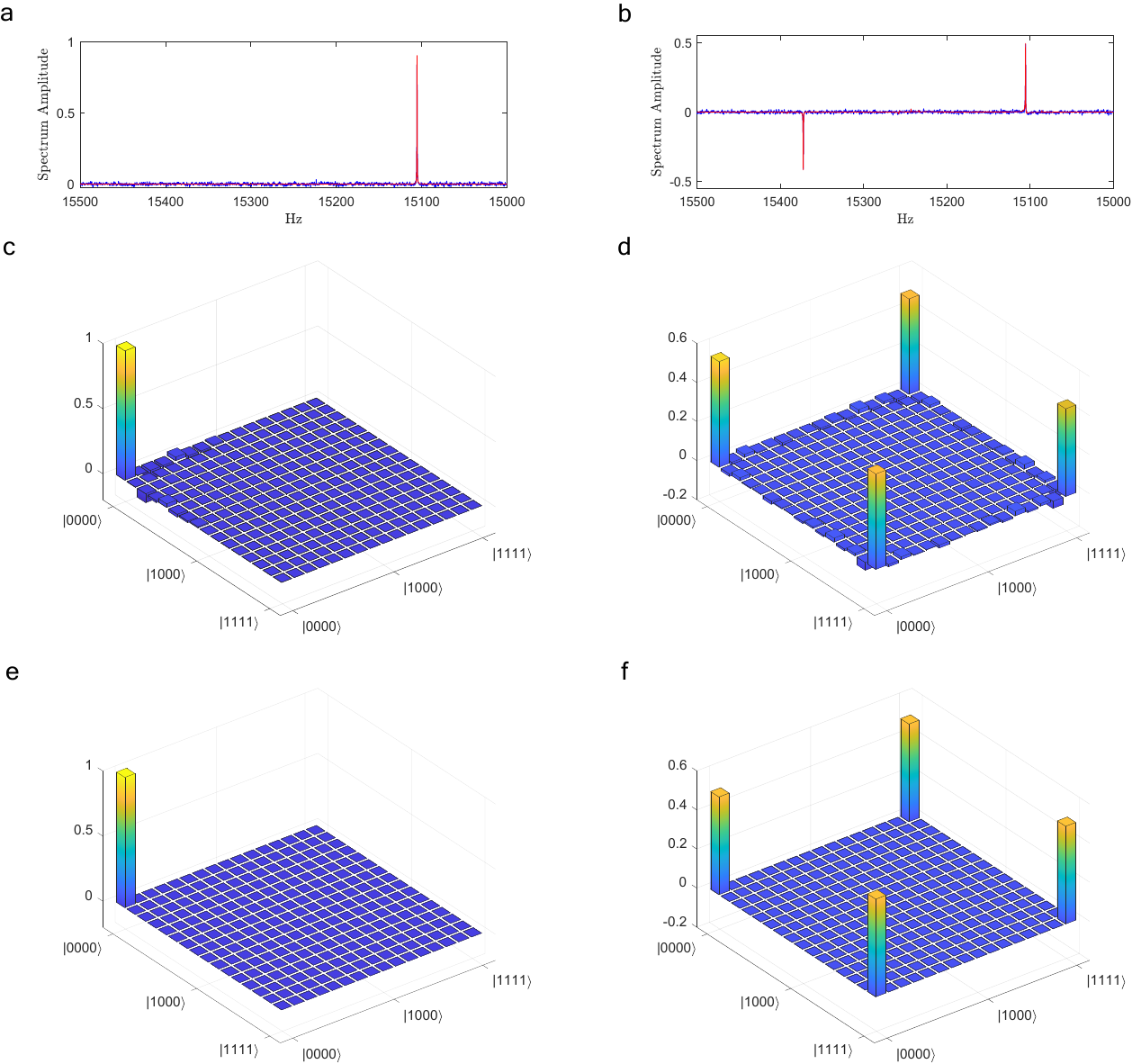}
    \caption{Experimental data and the corresponding tomography results.
    (a) Experimental $^{13}$C spectra of PPS $\rho_{0000}$ state.
    (b) Experimental $^{13}$C spectra of GHZ state with $R_{y}(\pi/2)$ rotation on $^{13}$C as the readout operator.
    (c) The PPS tomography and (d) the GHZ state tomography.
    (e) The ideal tomography of PPS state and (f) the ideal tomography of GHZ state.
    }
    \label{fig:exp}
\end{figure}

The spectrum of PPS $\rho_{0}$ is shown in Fig.\ref{fig:exp}(a), which is prepared from the thermal equilibrium state $\rho_{\mathrm{eq}}$ by shaped pulses.
The spectrum of GHZ state with $R_{y}(\pi/2)$ rotation on $^{13}$C as the readout operator is shown in Fig.\ref{fig:exp}(b), which is prepared by adding a pulse sequence generated by iGRAPE on PPS state.
We also perform full state tomography to reconstruct the density matrices of the PPS state as well as GHZ state.
In NMR setup, we first obtained only the deviation of the density matrix $\rho_{\Delta} = \rho - \mathbb{1}/2^{N}$ in NMR tomography, which can’t be regarded as a quantum state.
Therefore, we used the post-processing procedure to reach the density matrices in Fig.\ref{fig:exp}(c) (the PPS state) and Fig.\ref{fig:exp}(d) (the GHZ state) by introducing the constraints of the normalization $\tr(\rho) = 1$, the hermiticity $\rho = \rho^{\dagger}$ and the positive semi-definiteness $\rho \le 0$.
All of these constrains were realized by CVX toolbox in Matlab.
Therefore, the reconstructed matrix will satisfy all the requirements of the density matrix of a quantum state.
Fig.\ref{fig:exp}(e) and Fig.\ref{fig:exp}(f) shows the theoretical tomography of PPS and GHZ state.
The fidelity of a pure state $\ket{\psi}$ and a density matrix $\rho$ is defined as: $F(\ket{\psi},\rho) = \bra{\psi} \rho \ket{\psi}$.
The fidelity of reconstructed PPS state in Fig.\ref{fig:exp}(c) is $99.29\%$, and the fidelity of reconstructed GHZ state in Fig.\ref{fig:exp}(d) is $98.25\%$.

\section*{iGRAPE for superconducting quantum systems}

The superconducting quantum system we constructed here is a one-dimensional chain model with tunable coupling between qubits.
The parameters including idle frequency $\omega$ and anharmonicity $\eta$ are shown in Table \ref{tab:superconducting_para}.

\begin{table}[htbp]
    \centering
    \footnotesize
    \begin{tabular}{|c|c|c|c|c|c|c|c|c|c|c|c|c|}
    \hline
    & Q1 & Q2 & Q3 & Q4 & Q5 & Q6 & Q7 & Q8 & Q9 & Q10 & Q11 & Q12 \\
    \hline
    $\omega$ (GHz) & 4.978 & 4.183 & 5.192 & 4.352 & 5.110 & 4.226 & 5.030 & 4.300 & 5.142 & 4.140 & 4.996 & 4.260 \\
    \hline
    $\eta$ (MHz) & -248 & -204 & -246 & -203 & -247 & -202 & -246 & -203 & -244 & -203 & -246 & -201 \\
    \hline
    $T_{1}$ ($\mu$s) & 40.1 & 34.7 & 30.8 & 43.2 & 31.8 & 34.3 & 46.5 & 38.1 & 32.2 & 54.6 & 29.6 & 30.3 \\
    \hline
    $T_{2}^{*}$ ($\mu$s) & 7.9 & 1.5 & 6.3 & 2.4 & 4.9 & 2.7 & 6.8 & 2.3 & 5.1 & 3.5 & 5.9 & 3.0 \\
    \hline
    \end{tabular}
    \caption{Experimental parameters for all 12 qubits in the superconducting quantum system. $\omega$ is the idle frequency of qubit and $\eta$ is the anharmonicity. $T_{1}$, the energy relaxation time, and $T_{2}^{*}$, the dephasing time extracted from Ramsey experiment, are measured at idle frequency.}
    \label{tab:superconducting_para}
\end{table}

The duration $\Delta t$ for each propagator $\mathcal{U}_{k}^{[n,m]}$ in journal text is 0.05ns, and the parameters of shaped pulses are shown in Table \ref{tab:fig3}.

\begin{table}[H]
    \centering
    \footnotesize
    \begin{tabular}{|c|c|c|c|c|c|c|c|c|c|}
    \hline
    Number of qubits & 2 & 4 & 6 & 8 & 10 & 12 \\
    \hline
    iGRAPE pulse steps & 320, 300 & 380, 320, 300 & 420, 360, 320, 300 & 440, 380, 340, 300 & 460, 440, 360, 320, 300 & 500, 420, 360, 320, 300 \\
    \hline
    GRAPE pulse steps & 620 & 1000 & 1400 & 1460 & 1840 & 1900 \\
    \hline
    Transfer time (ns) & 31 & 50 & 70 & 73 & 92 & 95 \\
    \hline
    \end{tabular}
    \caption{Shaped pulses details of states preparation on superconducting quantum systems.}
    \label{tab:fig3}
\end{table}

\section*{The singular values of the entangled states}

The singular values of the quantum states generated by parameterized quantum circuit (see Fig.~4 in the journal text) are shown in Table \ref{tab:singular_5qubits} and Table \ref{tab:singular_8qubits}.

\begin{table}[htbp]
    \centering
    \footnotesize
    \begin{tabular}{|c|c|c|c|c|}
    \hline
    Layers & \multicolumn{4}{c|}{Singular values} \\
    \hline
    1 & 1. & 0. & 0. & 0. \\
    \hline
    3 & 0.919 & 0.394 & 0. & 0. \\
    \hline
    5 & 0.844 & 0.435 & 0.167 & 0.045 \\
    \hline
    7 & 0.736 & 0.615 & 0.228 & 0.173 \\
    \hline
    9 & 0.738 & 0.504 & 0.377 & 0.247 \\
    \hline
    \end{tabular}
    \caption{The singular values of 5-qubit quantum states in different layers of parameterized quantum circuits.}
    \label{tab:singular_5qubits}
\end{table}

\begin{table}[htbp]
    \centering
    \footnotesize
    \begin{tabular}{|c|c|c|c|c|c|c|c|c|c|c|c|c|c|c|c|c|}
    \hline
    Layers & \multicolumn{16}{c|}{Singular values} \\
    \hline
    1 & 1. & 0.003 & 0.003 & 0.002 & 0.002 & 0.001 & 0.001 & 0.001 & 0.001 & 0.001 & 0. & 0. & 0. & 0. & 0. & 0. \\
    \hline
    2 & 0.782 & 0.624 & 0.003 & 0.003 & 0.002 & 0.001 & 0.001 & 0.001 & 0.001 & 0.001 & 0.001 & 0. & 0. & 0. & 0. & 0. \\
    \hline
    3 & 0.782 & 0.623 & 0.003 & 0.002 & 0.002 & 0.002 & 0.002 & 0.002 & 0.001 & 0.001 & 0.001 & 0.001 & 0.001 & 0. & 0. & 0. \\
    \hline
    4 & 0.622 & 0.610 & 0.399 & 0.286 & 0.002 & 0.002 & 0.002 & 0.002 & 0.002 & 0.001 & 0.001 & 0.001 & 0.001 & 0.001 & 0. & 0. \\
    \hline
    5 & 0.662 & 0.547 & 0.418 & 0.298 & 0.003 & 0.002 & 0.002 & 0.002 & 0.002 & 0.001 & 0.001 & 0.001 & 0.001 & 0. & 0. & 0. \\
    \hline
    6 & 0.583 & 0.568 & 0.389 & 0.320 & 0.184 & 0.179 & 0.110 & 0.086 & 0.002 & 0.002 & 0.001 & 0.001 & 0.001 & 0.001 & 0. & 0. \\
    \hline
    7 & 0.575 & 0.521 & 0.417 & 0.316 & 0.254 & 0.181 & 0.145 & 0.082 & 0.002 & 0.002 & 0.001 & 0.001 & 0.001 & 0.001 & 0. & 0. \\
    \hline
    8 & 0.516 & 0.472 & 0.390 & 0.325 & 0.274 & 0.231 & 0.209 & 0.181 & 0.143 & 0.124 & 0.096 & 0.052 & 0.037 & 0.017 & 0.010 & 0.001 \\
    \hline
    9 & 0.508 & 0.467 & 0.393 & 0.341 & 0.277 & 0.238 & 0.212 & 0.171 & 0.143 & 0.115 & 0.079 & 0.057 & 0.039 & 0.028 & 0.017 & 0.010 \\
    \hline
    10 & 0.451 & 0.407 & 0.378 & 0.342 & 0.308 & 0.272 & 0.234 & 0.206 & 0.194 & 0.161 & 0.143 & 0.102 & 0.078 & 0.057 & 0.044 & 0.018 \\
    \hline
    11 & 0.457 & 0.409 & 0.359 & 0.338 & 0.313 & 0.276 & 0.249 & 0.216 & 0.189 & 0.150 & 0.132 & 0.104 & 0.085 & 0.059 & 0.022 & 0.004 \\
    \hline
    12 & 0.445 & 0.417 & 0.355 & 0.329 & 0.300 & 0.274 & 0.256 & 0.222 & 0.196 & 0.168 & 0.152 & 0.109 & 0.076 & 0.058 & 0.042 & 0.014 \\
    \hline
    \end{tabular}
    \caption{The singular values of 8-qubit quantum states in different layers of parameterized quantum circuits.}
    \label{tab:singular_8qubits}
\end{table}